\documentclass[twocolumn,showpacs,amssymb,preprintnumbers,prl,floatfix]{revtex4}
\usepackage{rotating,epsfig,dcolumn}
\usepackage{graphicx,bm}

\begin{document}
\title{The superdeformed excited band of $^{40}$Ca}
\author{E. Caurier $^a$, F. Nowacki $^a$,
   A.~Poves $^b$ and A. Zuker  $^a$}
\affiliation{
(a) IReS, B\^at27, IN2P3-CNRS/Universit\'e Louis
Pasteur BP 28, F-67037 Strasbourg Cedex 2, France\\
(b) Departamento de Fisica Te\'orica, C-XI. Universidad Aut\'onoma
de Madrid, E-28049, Madrid, Spain}
\date{\today}
\begin{abstract}
  The superdeformed band, recently discovered in   $^{40}$Ca is
  analysed in an spherical shell model context. Two major oscillator
  shells, $sd$ and $pf$ are necessary to describe it.
  The yrast band of the fixed 8p-8h
  configuration fits extremely well with the experimental energies and
  transition rates of the superdeformed band. The 4p-4h
  configuration generates a normally deformed band plus a
  $\gamma$-band pattern which are also present in the experimental data.

\end{abstract}
\maketitle

 The existence of excited deformed bands in spherical nuclei is a well
 documented fact, dating back to the 60's. A  classical example is
 provided by the four particle four holes and eight particles eight
 holes states in $^{16}$O, starting at
 6.05~MeV and 16.75~MeV of excitation energy \cite{O16a,O16b}.
 However, it is only recently that
 similar bands, of deformed and even superdeformed character, have been
 discovered in other medium-light nuclei such as
 $^{56}$Ni~\cite{ni56exp}, 
 $^{36}$Ar~\cite{ar36exp} and  $^{40}$Ca~\cite{ca40exp} and explored
 up to high spin.
 These experiments have been possible thanks to the advent of large
 arrays of $\gamma$ detectors, like Gammasphere or Euroball. 
 One characteristic feature of these bands is that they belong to
 rather well defined spherical shell model configurations; for
 instance, the deformed excited band in $^{56}$Ni can be associated with
 the configuration (1f$_{7/2}$)$^{12}$ (2p$_{3/2}$, 1f$_{5/2}$,
 2p$_{1/2}$)$^{4}$ while the (super)deformed band in  $^{36}$Ar has
 the structure $(sd)^{16}$ $(pf)^{4}$. While many approaches are
 available for the microscopic description of these bands (Cranked
 Nilsson-Strutinsky, Cranked Hartree-Fock Bogolyuvov, etc.) the
 interacting shell model is, when affordable, the prime choice. The
 mean field approach to N=Z nuclei, has 
 problems related to the proper treatment of the
 proton-neutron pairing in its isovector and isoscalar channels. On
 the shell model side, the problems come from the size of the valence
 spaces needed to acomodate the np-nh configurations, from the danger
 of center of mass contamination and from the occurrence of high level
 densities in the energy region where some of the members of
 the excited bands lay.

 Prompted by our previous experience in the description of such states in
 $^{56}$Ni and $^{36}$Ar we have undertaken to explain the occurrence
 of deformed and superdeformed bands in $^{40}$Ca as reported very
 recently by Gammasphere experimenters \cite{ca40exp}. The states we
 aim to are dominantly core excitations from the $sd$-shell to the
 $pf$-shell. The natural valence space would thus comprise both major
 oscillator shells. However, the inclusion of the 1d$_{5/2}$ orbit in
 the valence space produces a 
 huge increase in the size of the
 basis and massive center of mass effects, thus we were 
 forced to exclude it from the valence
 space already in the  $^{36}$Ar calculations and we will proceed
 equally now, taking a closed core of $^{28}$Si. We are aware that this
 truncation will somehow reduce, although not drastically,  the quadrupole
 coherence of our solutions. Hence, our valence space will consist of  the
 orbits  2s$_{1/2}$, 1d$_{3/2}$, 1f$_{7/2}$, 2p$_{3/2}$, 1f$_{5/2}$,
 2p$_{1/2}$. The effective interaction is the same used in
 ref.~\cite{ar36exp} denoted  $sdpf.sm$.
 In a first step, we keep fixed the number of particles excited from
 the $sd$ to the $pf$ orbits. We surmise that the 4p-4h configurations
 are responsible for the band built on the first excited 0$^+$
 state of   $^{40}$Ca at 3.352 MeV excitation energy and for some
 related states, while the superdeformed band  pertains to the 8p-8h
 configurations. Indeed, the assumption of a fixed np-nh character for the
 states is not fully realistic, some mixing being always present.
 In spite of this, we believe that the np-nh states contain most of the
 physics relevant to the problem. In a second step we will deal
 explicitely with some features of the mixing.

\begin{table}[h]
\caption{{\label{tab:4p4h.energy}}
 Calculated $\gamma$ energies for the 4p-4h configuration's yrast and
 $\gamma$ bands in $^{40}$Ca, 
  compared to the experimental bands labeled 2 and 4 in \cite{ca40exp}
 (energies in keV)}
 \begin{ruledtabular}
\begin{tabular}{rrrrrr}
\multicolumn{2}{l}{E$_{\gamma}$(J$_{i}$ $\rightarrow$ J$_{f}$)} 
    & band 2& th. yrast  
 & band 4 & th. $\gamma$-band \\   
\hline
 2  & 0 & 553   &772   &    &   \\
 3  & 2 &    &   & 781   & 819  \\
 4  & 2 &1375    & 1121  & 1260   & 1244  \\
 5  & 4 &    &   & 1369   & 1187  \\
 6  & 4 &1653    &1458   &    &   \\
 7  & 5 &    &   &1538    & 1501  \\
 8  & 6 &2374    &2210   &   &   \\
 9  & 7 &    &   & 2773   & 2346  \\
10  & 8 & 2381   & 2050  &    &   \\
11  & 9 &    &   & 1827   &1518   \\
12  & 10 & 2546   & 2866  &    &   \\
13  & 11 &    &   & 3044   & 1943  \\
14  & 12 & 2297    & 1923  &    &   \\
16  & 14 & 4050   & 3374  &    &   \\
\end{tabular} 
\end{ruledtabular}
\end{table}

 The $\gamma$-ray energies obtained in the diagonalizations in the
 4p-4h space  are gathered
 in table~\ref{tab:4p4h.energy} and compared with the experimental 
 results from 
 \cite{ca40exp}. The agreement is satisfactory. The most salient
 aspect of the calculated results is
 the triaxial character of the solution, with a  well developped
 $\gamma$-band based in the second 2$^+$ state of the 4p-4h
 configuration. The yrast sequence follows nicely the 
 pattern of the experiment  (Band 2 in \cite{ca40exp}) within 200 keV, 
 except for the highest spin member of the
 band, J=16$^+$, that comes out clearly too low. The experimental band
 presents an upbending at  J=10$^+$, that the calculation turns out
 into a slight backbending. The calculated $\gamma$-band can be put
 in correspondence with the experimental states grouped in Band 4.

\begin{table}[h]
\caption{{\label{tab:4p4h.qyrast}}
Quadrupole properties of the 4p-4h configuration's yrast-band in
 $^{40}$Ca (in e$^2$fm$^4$ and efm$^2$)}
 \begin{ruledtabular}
\begin{tabular}{ccccc}
 J & B(E2)(J $\rightarrow$ J-2) & Q$_{spec}$ & Q$_0$(s) & Q$_0$(t) \\   
\hline
 2 & 266 & -28.6 & 100 & 116 \\
 4 & 356 & -39.9 & 110 & 112 \\
 6 & 328 & -47.5 & 119 & 102 \\
 8 & 298 & -42.3 & 101 &  96 \\
10 & 220 & -44.5 & 102 &  81  \\
12 &  80 & -38.6 &  87 &  94 \\
14 & 111 & -40.9 &  90 &  57  \\
16 &  46 & -34.7 &  76 &  37 \\
\end{tabular} 
\end{ruledtabular}
\end{table}

 The B(E2)'s and the quadrupole moments
 corresponding to the (4p-4h)-yrast states are presented in 
 table~\ref{tab:4p4h.qyrast}. The calculations assume harmonic
 oscillator single particle wave functions with size parameter
 $b$=1.97~fm, extracted from the experimental charge radius of
 $^{40}$Ca. Effective charges $\delta$e$_{\pi}$=$\delta$e$_{\nu}$=0.5
 are used.
   The intrinsic quadrupole moments behave quite regularly up to the
 backbending regime; from there on, there is not a real intrinsic state.
 This is manifest in the rapid decrease of the values
 Q$_0$(t) (the intrinsic quadrupole moment extracted from the B(E2)'s
 assuming K=0) and in the large difference between  Q$_0$(t) and  Q$_0$(s)
 (the  intrinsic quadrupole moment extracted from the spectroscopic
 quadrupole moment, assuming K=0). In the experimental article, a
 value  Q$_0$(t)=0.74$\pm$0.14~eb is obtained from the measured
 fractional Doppler shifts. Our values are somewhat larger but 
 in reasonable  agreement with this
 figure, that corresponds to $\beta$=0.27, a typical value for a
 normally deformed band.
 The quadrupole properties of the $\gamma$-band are gathered in
 table~\ref{tab:4p4h.qgamma}. They are consistent with our picture of an
 triaxial rotor. Assuming K=2, we can extract also the intrinsic
 quadrupole moments, that keep approximately constant along the
 band. The oscillations are much larger than before reflecting a less
 well defined intrinsic state. Very
 strong $\Delta$J=1 E2 transitions are calculated between the lower
 states of the band. At higher spins, the $\Delta$J=2 transitions take
 over rapidly.

\begin{table}[h]
\caption{{\label{tab:4p4h.qgamma}}
Quadrupole properties of the 4p-4h configuration's $\gamma$-band in
 $^{40}$Ca (in e$^2$fm$^4$ and efm$^2$)}
 \begin{ruledtabular}
\begin{tabular}{ccccc}
 J & B(E2)(J $\rightarrow$ J-1) & B(E2)(J $\rightarrow$ J-2) & Q$_{spec}$ & Q$_0$(s)  \\   
\hline
 2 &     &     &  26.8 &  93 \\
 3 & 411 & 132 &  -0.3 &   - \\ 
 4 & 440 & 210 & -16.8 & 116  \\ 
 5 & 229 & 189 & -25.4 & 110 \\ 
 6 & 177 & 189 & -35.7 & 125 \\ 
 7 & 123 & 229 & -35.7 & 110 \\
 8 &  50 & 189 & -46.7 & 134 \\ 
 9 &  82 & 200 & -35.3 &  95 \\
10 &  43 & 169 & -42.6 & 110 \\ 
11 &  26 & 113 & -46.3 & 116   \\
12 &  29 &   6 & -38.6 &  95 \\ 
13 &   9 &  83 & -40.9 &  98 \\

\end{tabular} 
\end{ruledtabular}
\end{table}

\begin{figure}[h]
  \begin{center}
    \leavevmode 
    \epsfig{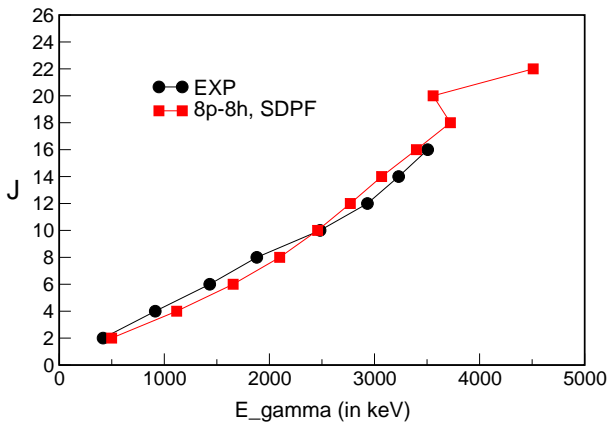}
     \caption{The $^{40}$Ca superdeformed band, experiment
    vs. the 8p-8h calculation in the sdpf valence space}
    \label{fig:8p-8h}
  \end{center}
\end{figure}

The natural continuation of this study would have been  to compute the yrast
band of the 8p-8h configuration in the same space and with the same
interaction, aiming to the explanation of the experimental
superdeformed band. However, the extremely large dimension of the
basis ($\approx$~10$^9$) poses a serious problem. These dimensions are
at present tractable with the new versions of the codes ANTOINE and
NATHAN  \cite{antoine}, but, to calculate the complete band will demand a
huge computational effort.
Nevertheless, our experience in the $pf$-shell tells us that, to
limit the maximum number of particles in the 1f$_{5/2}$ and
 2p$_{1/2}$ to two, produces results that are already very close to
 the complete ones. The extra one-particle excitations contribute
 mainly to the quadrupole coherence while, the two-particle
 excitations improve the pairing content of the solutions.
 These are huge calculations too (dim
 $\approx$ 3x~10$^8$) but nonetheless doable.
 We have carried out the calculations in the fixed 8p-8h configuration
 using the interaction $sdpf.sm$. The gamma-ray energies along the yrast
 sequence   are compared
 to the experimental results (Band 1 in ref.~\cite{ca40exp})
 in fig.~\ref{fig:8p-8h} in the form of a backbending plot.
 The interaction reproduces very  satisfactorily the experimental
 results. The only difference is the change of slope in the experimental
 curve at J=10, not reproduced by the calculation, that is probably an
 effect of the mixing.
 Notice that the experimental (and the predicted)
 bands are very regular, showing  no backbending, contrary to
 the situation in $^{48}$Cr \cite{cr48}. Due to the presence of $sd$
 particles in pseudo-SU3 orbitals, the alignment process is pushed to
 higher angular momentum. In fact our calculations predict a
 slight backbending at J=20 and an unfavoured band termination at J=22,24.

In table~\ref{tab:8p8h.sdpf} we have collected the quadrupole
properties.  The  Q$_0$'s extracted from the B(E2)'s and from the
spectroscopic quadrupole moments are nearly
equal and  reasonably constant up to the backbending,
 supporting the existence of
a robust intrinsic state. The calculated Q$_0$ of 172~efm$^2$ is
in perfect agreement with the experimental value
Q$_0$(t)=1.80$^{+0.39}_{-0.29}$, obtained from the fractional Doppler
shifts. This experimental value corresponds to a deformation
$\beta$$\sim$0.6, i.e. to a superdeformed shape.  The 
calculation predicts a slight decrease of the deformation with
increasing J, while
experimentally it seems to remain constant until the highest measured spin
state (J=16). This departure may be due to the blocking of the 1d$_{5/2}$ orbital.
Despite that, the calculation at fix
particle-hole number contains  most of the relevant physics
of the superdeformed band in $^{40}$Ca. The deformation that our
calculation produces is probably the highest ever obtained in a shell
model calculation describing a ``bona fide'' rotational band. As a
matter of fact it almost saturates the SU(3) limit of the intrinsic
quadrupole moment in these two major shells; Q$_0$=190~efm$^2$, or the
--more realistic-- quasi-SU(3) one; Q$_0$=180~efm$^2$. At the
band termination, the B(E2)'s drop to zero while the
spectroscopic quadrupole moments keep constant, reflecting the
transition from the collective to the aligned regime.

\begin{table}[h]
\caption{{\label{tab:8p8h.sdpf}}
Quadrupole properties of the 8p-8h configuration's yrast-band in $^{40}$Ca,
calculated in the $sdpf$ valence space}
 \begin{ruledtabular} 
\begin{tabular}{rccrr}
 J & B(E2)(J $\rightarrow$ J-2)&
 Q$_{spec}$ &  Q$_0$(t) & Q$_0$(s) \\
\hline
 2 & 589 &  -49.3 &  172 & 172 \\
 4 & 819 &  -62.4 &  170 & 172 \\
 6 & 869 &  -68.2 &  167 & 171 \\
 8 & 860 &  -70.9 &  162 & 168 \\
10 & 823 &  -71.6 &  157 & 164 \\
12 & 760 &  -71.3 &  160 & 160 \\
14 & 677 &  -71.1 &  149 & 157 \\
16 & 572 &  -72.2 &  128 & 158 \\
18 & 432 &  -75.0 &  111 & 162  \\
20 &  72 &  -85.1 & & \\
22 &   8 &  -79.1 & & \\
24 &   7 &  -81.5 & & \\
\end{tabular} 
\end{ruledtabular}
\end{table}

\begin{figure}[h]
  \begin{center}
    \leavevmode 
    \epsfig{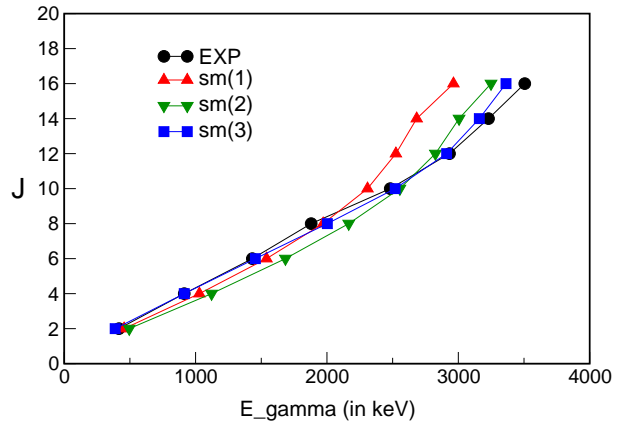}
    \caption{The $^{40}$Ca superdeformed band, experiment
    vs. different 8p-8h calculations in the zbm2 valence  space}
    \label{fig:zbm2}
  \end{center}
\end{figure}

The next step is the study of the mixing of the different np-nh
configurations. Unluckily, in this valence space, this goal is definitely
beyond our computational possibilities --even with the truncation
adopted above-- because it demands the calculation of many states of
the same total angular momentum.  We try to circumvent these
limitations by reducing once more the valence space, eliminating
completely the upper $pf$-shell orbits.  The active orbits will then
be 2s$_{1/2}$, 1d$_{3/2}$, 1f$_{7/2}$, 2p$_{3/2}$.  This valence space
will be called zbm2, to emphasize its links with the space
(1p$_{1/2}$, 1d$_{5/2}$, 2s$_{1/2}$) used in the late 60's to describe
the core excited states of $^{16}$O \cite{zbm}.  We have tried to
evaluate the effects of this further truncation by comparing the 8p-8h
results in the two spaces, with the interaction $sdpf.sm$.  The
the new $\gamma$-energies are presented in fig.~\ref{fig:zbm2}
 under the label sm(1).

 The yrast band 
behaves nicely for the lower spins, but is too compressed beyond J=10.
The quadrupole collectivity turns out to be  a 10$\%$ smaller and
decreases more rapidly at high spins than it does in the fuller calculation.
This suggest that an increase of the quadrupole-quadrupole interaction
in the $pf$ sector of the interaction could mock the full space results.
Therefore, we have increased a 10$\%$ the $qq$ interaction and
repeated the calculations in the zbm2 space. The results for the
energies are plotted in fig.~\ref{fig:zbm2} (label sm(2)), and we see
that the band has now a much better behaviour and the correct span.
However, the quadrupole properties remain basically unchanged. 
 We have tried
other interactions and/or renormalizations and our conclusion is that
any decent interaction produces the same intrinsic
state i.e. the quadrupole collectivity has reached saturation
in this valence space. 
 In order to increase the deformation and to
keep it constant for the higher  spins one must
enlarge the valence
space, this will increase the intrinsic quadrupole moment a 10$\%$ upon
inclusion of the upper $pf$-shell and another 10$\%$ upon opening the
1d$_{5/2}$ orbital.
 Thus, if we keep the usual effective charges, we have to be  
 aware that the quadrupole
coherence in the zbm2 space will always be a 20$\%$ smaller than in
the full space of the two major shells.

A new interaction  has been built recently, specifically for
 the zbm2  valence space, and
used to describe the radii isotope shifts in the Calcium isotopes
\cite{isoca}. It was based  on $sdpf.sm$ with  mostly monopole changes.
Following our discussion above, we have
increased a ten percent the quadrupole-quadrupole interaction
 in the $pf$ sector of the zbm2 space. In addition, 
in~\cite{isoca} an off-diagonal, cross-shell schematic isovector 
pairing was substracted
from the initial interaction to avoid double counting. Here we choose
to reduce all the off-diagonal cross-shell matrix elements a 25$\%$.
It is with this interaction that we shall proceed to study the
 mixing. The results for the fixed 8p-8h configuration can be found
 also in fig.~\ref{fig:zbm2} with the label sm(3). As expected both
 the energies and the quadrupole properties are very
 close to the sm(2) set.

\begin{figure}[h]
  \begin{center}
    \leavevmode 
    \epsfig{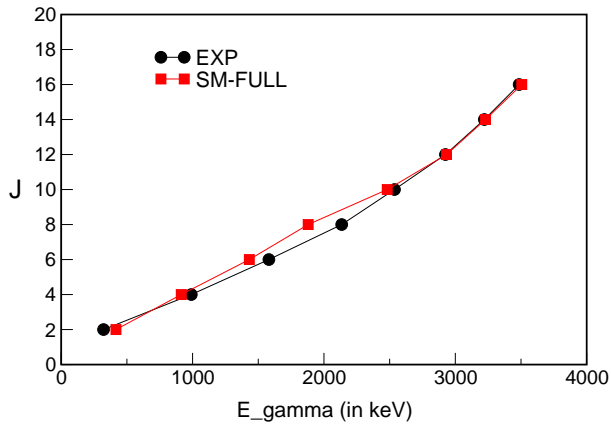}
    \caption{The $^{40}$Ca superdeformed band, experiment
    vs. mixed calculation in the full zbm2 space}
    \label{fig:sdfull}
  \end{center}
\end{figure}

 The final step consists in finding the mixing among the different
 np-nh configurations. This is a formidable task because the np-nh
 series is not easy to truncate. In earlier studies \cite{np-nh} we
 have reached the conclusion that, in order to have the correct mixing
 in  a given np-nh configuration, the valence space must contain at
 least all the configurations up to 
 (np+4)-(nh+4). This is easy to understand, because 
 the pairing interaction mixes configurations  that
 are 2p-2h apart. But the amount of mixing depends also of their
 relative possition, therefore, if the configurations
 that are 4p-4h apart are not included, the 2p-2h ones will be too
 high and will not renormalize the reference states properly. 
 That's why the zbm2  valence space cannot provide
 the proper renormalization to the 8p-8h band, because the 10p-10h and 
 12p-12h configurations cannot develop enough collectivity due to the
 closure of the 1d5/2 orbit. This has two effects; first, as these
 configurations lay much too high in energy, they do not mix enough
 with the 8p-8h ones and the reduced gain in energy has to be
 compensated with an artificial lowering of the 8p-8h
 configurations. But even doing that, the absence of mixing with other
 configurations that have similar of larger quadrupole contents, makes
 the mixed results unphysically less collective than the  unmixed ones.
 With all these caveats we have tried to get a first glimpse
 on the mixed results, that we comment now briefly. We have played
 freely with the $sd$-$pf$ global monopoles in order to locate the
 three lower 0$^+$ states of $^{40}$Ca close to their experimental
 values. The structure of the ground state is the expected one; about
 60$\%$ closed shell, with mainly 2p-2h mixing. The first excited 
 0$^+$ is dominantly  (60$\%$) 4p-4h with mainly 6p-6h and 8p-8h
 mixing. It corresponds to the bandhead of the experimental Band
 2. The third one is 60$\%$ 8p-8h, and the mixing is
 dominantly 4p-4h. The 10p-10h and 12p-12h component are --as we had
 anticipated-- completely absent.
 The calculation of the three lower 0$^+$ is straightforward, however,
 the calculation of the excited states belonging to the 4p-4h or the
 8p-8h bands is not, because these are most often  drowned in a sea of
 other uninteresting states. To overcome this difficulty we 
 select as starting vectors in the Lanczos procedure
 the eigenstates of the band obtained in the np-nh space. This choice
 accelerates the
 convergence of the states we are seeking and makes it possible to keep
 track of them in case of fragmentation. We have used this method to 
 obtain the mixed  superdeformed band, starting with the 8p-8h states.  
 The excitation
 energies change little, but enough to improve the quality of
 the agreement with
 the experimental data of the 8p-8h calculation.
 (see figure~\ref{fig:sdfull}).
 However, the quadrupole moments and transition probabilities get
 eroded --30$\%$ to 50$\%$-- by the mixing or even completely washed
 out at the top  of
 the band. As we have discussed above, this is  an intrinsic
 limitation of the valence space that we cannot ovecome.

 In conclusion we have performed a study of the many particle many
 hole configurations in  $^{40}$Ca, aiming to understand its recently
 discovered superdeformed excited band. We have shown that the yrast
 band of the 8p-8h
 configuration in the $sd$-$pf$ valence space reproduces
 very well the experimental results and represents a show-case of
 superdeformed band described by the spherical shell model.

This work is partly supported by the IN2P3(France) CICyT(Spain)
collaboration agreements. AP's work is supported by  MCyT (Spain),
 grant BFM2000-30.

\end{document}